\documentclass[twocolumn,showpacs,preprintnumbers,amsmath,amssymb]{revtex4}

\usepackage{graphicx}
\usepackage{dcolumn}
\usepackage{bm}

\begin{document}

\preprint{APS/123-QED}

\title{Li$_{1.1}$(Zn$_{1-x}$Cr$_x$)As: Cr doped I-II-V Diluted Magnetic Semiconductors in Bulk Form}

\author{Quan Wang$^{1}$, Huiyuan Man$^{1}$, Cui Ding$^{1}$, Xin Gong$^{1}$, Shengli Guo$^{1}$, Huike Jin$^{1}$, Hangdong Wang$^{2}$, Bin Chen$^{2}$, F.L.
Ning$^{1,}$}\email{ningfl@zju.edu.cn}

\affiliation{$^{1}$Department of Physics, Zhejiang University,
Hangzhou 310027, China} \affiliation{$^{2}$Department of Physics,
Hangzhou Normal University, Hangzhou 310016, China}

\date{\today}


\begin{abstract}
We report the synthesis and characterization of bulk form diluted
magnetic semiconductors I-II-V  Li$_{1.1}$(Zn$_{1-x}$Cr$_x$)As ($x$
= 0.03, 0.05, 0.10, 0.15) with a cubic crystal structure identical
to that of III-V GaAs and II-VI zinc-blende ZnSe. With $p$-type
carriers created by excess Li, 10\% Cr substitution for Zn results
in a ferromagnetic ordering below $T_C$ $\sim$ 218 K. Li(Zn,Cr)As
represents another magnetic semiconducting system with the advantage
of decoupling carriers and spins, where carriers are created by
adding extra Li and spins are introduced by Cr substitution for Zn.
\end{abstract}

\pacs{75.50.Pp, 71.55.Ht, 76.75.+i}

\maketitle


The observation of ferromagnetic ordering in III-V (Ga,Mn)As
\cite{Ohno} thin-film has generated extensive research into diluted
magnetic semiconductors (DMS) \cite{Jungwirth,Dietl1,Zutic}. The
highest Curie temperature, $T_C$, has been reported as $\sim$200 K
with Mn doping levels of $\sim$12 $\%$ in (Ga,Mn)As
\cite{ZhaoJH1,ZhaoJH2}. The application of spintronics may become
possible once $T_C$ reaches room temperature \cite{Zutic}. However,
the research is hindered by two inherent difficulties: one is that
the mismatch of valences of Ga$^{3+}$ and Mn$^{2+}$ prohibits the
fabrication of bulk form specimens with higher Mn doping levels; the
other is that it is difficult to determine precisely the amount of
Mn that substitutes Ga, which donates a hole and acts as a local
moment, since some Mn impurities enter interstitial sites
\cite{Jungwirth}. In the II-VI family of DMS, i.e.,
(Zn$_{1-x}$Mn$_x$)Se, the chemical solubility can be as high as 70\%
and bulk form specimens are available \cite{Pajaczkowska,Furdyna}.
The isovalent substitution of Mn for Zn, however, makes it difficult
to control the carrier density \cite{Wojtowicz,Morkoc}, which is as
low as 10$^{17}$ cm$^{-3}$,  and the magnetic moment size is as
small as 0.01 $\mu_B$/Mn \cite{Furdyna,Shand}.

To overcome the difficulties encountered in III-V and II-VI DMS,
Masek et al proposed that I-II-V direct-gap semiconductor LiZnAs may
be a good candidate for fabrication of next generation of DMS
\cite{Masek}. LiZnAs is a direct gap semiconductor with a band gap
of 1.6 eV \cite{Bacewicz}. It has a cubic structure, similar to that
of zinc-blende GaAs and ZnSe, as shown in Fig. 1. More
interestingly, if we view the combination of (Li$^{1+}$Zn$^{2+}$) as
Ga$^{3+}$, (LiZn)As becomes GaAs; alternatively, if we view the
combination of (Li$^{1+}$As$^{3-}$) as Se$^{2-}$, Zn(LiAs) becomes
ZnSe. The I-II-V semiconductor LiZnAs shares some common
characteristics with both III-V and II-VI semiconductors, and has
two superior advantages from the view of synthesis: (1) the
isovalent substitution of Mn for Zn overcomes the small chemical
solubility; (2) the carrier density can be controlled by
off-stoichiometry of Li concentrations.

Recently, Deng et al. successfully synthesized two bulk I-II-V DMS
systems, Li(Zn,Mn)As \cite{Deng1} and Li(Zn,Mn)P \cite{Deng2}, with
$T_C$ $\sim$ 50 K. The availability of the bulk form specimens
readily enables the microscopic investigation of DMSs based on
typical bulk probes of magnetism, such as nuclear magnetic resonance
(NMR), muon spin relaxation ($\mu$SR) and neutron scattering, which
would provide unique microscopic information of the ferromagnetism.
For example, through the NMR measurement of local static and dynamic
susceptibility at Li sites, Ding et al found that Mn atoms are
homogeneously doped in Li(Zn,Mn)P, and Mn spin-spin interactions
extend over many unit cells, which explains why DMSs could exhibit a
relatively high $T_C$ with such a low density of Mn \cite{Ding2}.

\begin{figure}
\includegraphics[width=8cm]{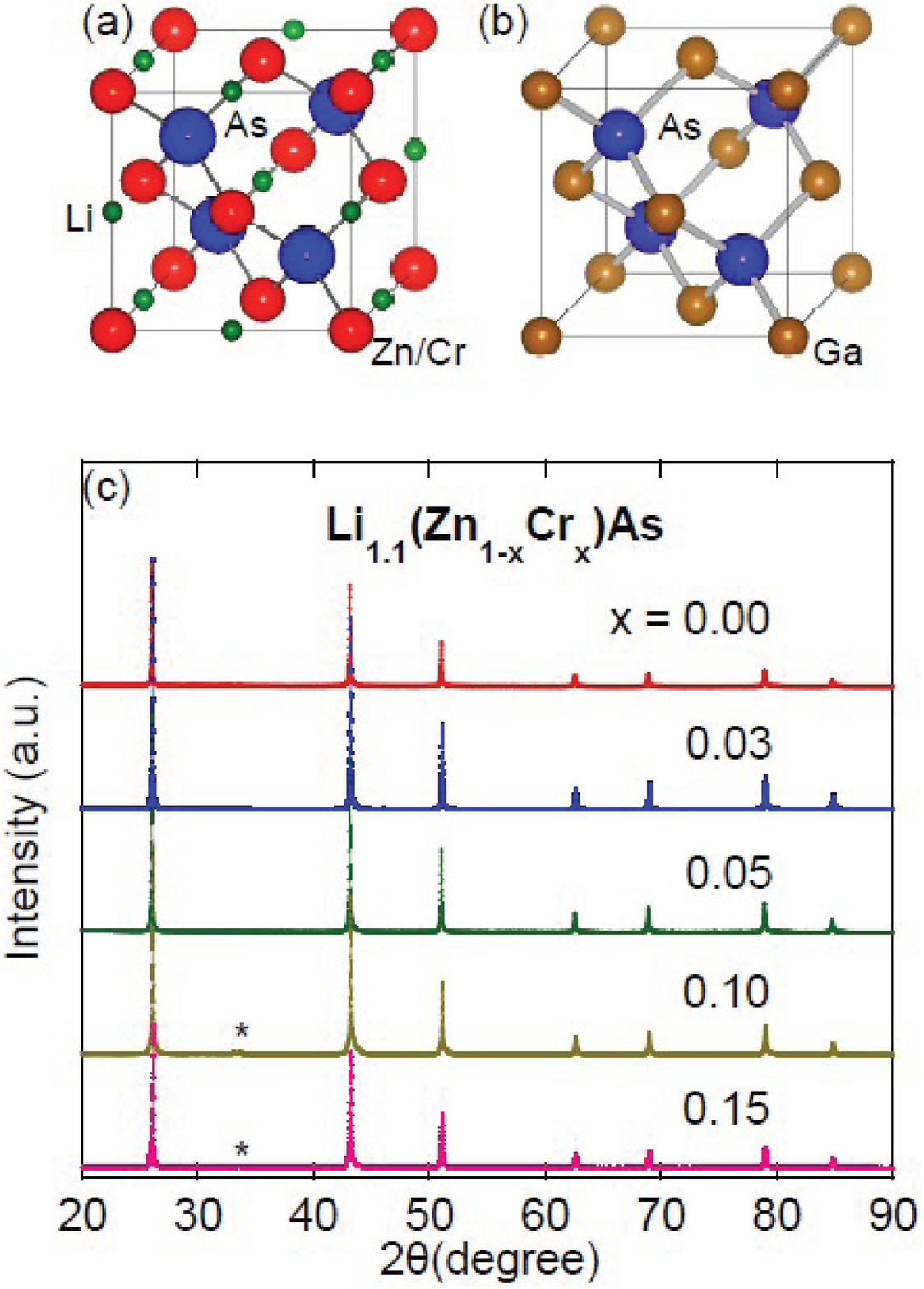}
\caption{\label{Fig1:epsart} (Color online). The crystal structure
of (a) Li(Zn,Cr)As, and (b) GaAs. (c) X-ray diffraction pattern of
Li$_{1.1}$(Zn$_{1-x}$Cr$_x$)As. Traces of impurity CrAs ($\ast$) are
marked for $x$ $\geq$ 0.1.}
\end{figure}
Very recently, three other bulk DMS systems with decoupling of
carriers and spins have been reported. Ding et al reported the
successful fabrication of a new bulk ``1111" DMS material,
(La$_{1-x}$Ba$_x$)(Zn$_{1-x}$Mn$_x$)AsO, with $T_C$ up to 40 K
\cite{Ding1}; Yang et al reported the ferromagnetic ordering at 210
K for Sr and Mn doped LaOCuS system \cite{Yangxj}; Zhao et al.
reported another ferromagnetic DMS system,
(Ba,K)(Zn,Mn)$_{2}$As$_{2}$ with $T_C$ up to $\sim$180 K
\cite{ZhaoK}. The Curie temperature of the latter two bulk DMSs is
already comparable to that of thin film (Ga,Mn)As \cite{ZhaoJH1,
ZhaoJH2}.

In addition to Mn, magnetic Cr atoms have also been extensively
doped into various III-V and II-VI semiconductors for fabrication of
DMSs. For example, room temperature ferromagnetism has been observed
in (Zn,Cr)Te thin films \cite{Saito1}. On the other hand, the Curie
temperature $T_C$ is relatively low, i.e., $\sim$ 10 K in Cr doped
GaAs \cite{Saito2, ZhaoJH3, ZhaoJH4}. The reason for such different
$T_C$ values between III-V and II-VI DMSs is still largely unknown.
Furthermore, the electron configuration of Cr atom is
(3d)$^5$(4s)$^1$, leaving the controversy about the valence of Cr
and its spin state \cite{Kuroda,ZhaoJH3, ZhaoJH4, Saito2}. This
situation is in contrast with the case of Mn$^{2+}$ whose ``+2"
valence and high spin state have been widely accepted.

In this letter, we report the synthesis and characterization of Cr
doped Li(Zn,Cr)As semiconductors. The carriers are introduced by
excess 10$\%$ Li, and the spins are introduced by the substitution
of Cr for Zn, respectively. We found that 3$\%$ Cr doping induces
local moments, but no ferromagnetic ordering is formed. At the
doping level of 5$\%$, a ferromagnetic ordering is developed, as
evidenced by a sudden enhancement of magnetization, and the
parallelogram-shaped hysteresis loops below $T_C$ $\sim$ 106 K.
$T_C$ is maximized to $\sim$ 218 K with 10$\%$ Cr doping, and
decreases to 201 K at the doping level of 15$\%$.

\begin{figure}
\includegraphics[width=8cm]{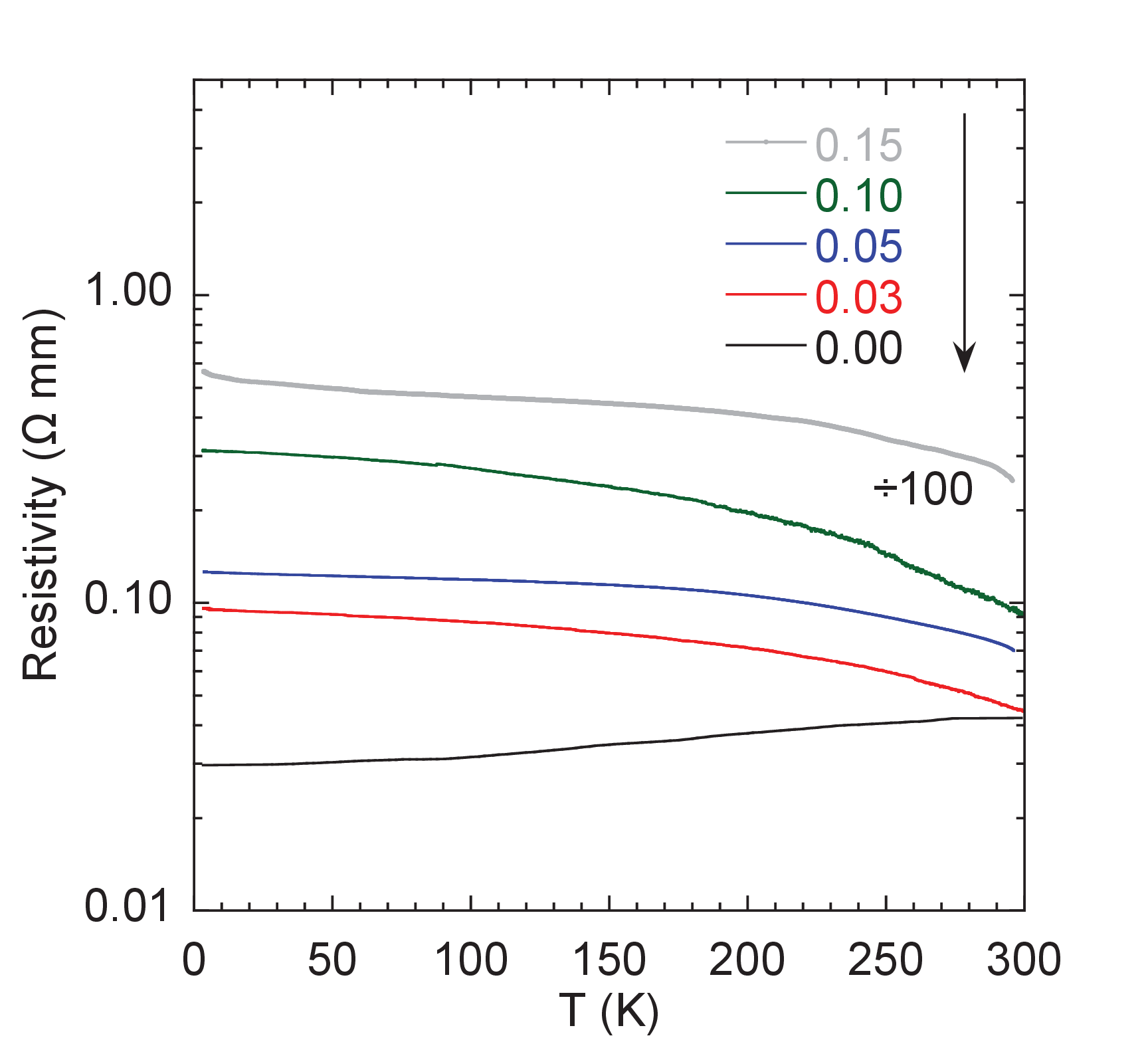}
\caption{\label{Fig2:epsart} (Color online) The electrical
resistivity of Li$_{1.1}$(Zn$_{1-x}$Cr$_x$)As with $x$ = 0, 0.03,
0.05, 0.10, 0.15; note that the magnitude of $x$ = 0.15 is divided
by 100 to fit into the window, and the resistivity of $x$ = 0 is
reproduced from ref. \cite{Deng1}.}
\end{figure}

We synthesized the polycrystalline specimens
Li$_{1.1}$Zn$_{1-x}$Cr$_{x}$As ($x$ = 0.03, 0.05, 0.10, 0.15) by the
solid state reaction method. We mixed the elements of Li (99.9\%),
Zn (99.9\%), Cr (99.99\%) and As (99\%) and slowly heated the
mixture to 500$^{\circ}$C in evacuated silica tubes, and held for 60
hours before cooling down to room temperature at the rate of
50$^{\circ}$C/h. The product was further pressed into pellet, heated
to 680 $^{\circ}$C, and held for 60 more hours. The polycrystals
were characterized by X-ray diffraction at room temperature and dc
magnetization by Quantum Design SQUID. The electrical resistance was
measured on sintered pellets with typical four-probe method.

We show the crystal structure of Li$_{1.1}$Zn$_{1-x}$Cr$_{x}$As and
the X-ray diffraction patterns in Fig. 1. Bragg peaks from the
parent compound LiZnAs can be well indexed by a cubic structure. The
single phase is conserved with the doping level up to $x$ = 0.05.
Small traces of CrAs (an antiferromagnet with $T_N$ $\sim$ 260 K
\cite{CrAs1,CrAs2}) impurities appear at $x$ = 0.10 and 0.15, as
marked by the stars. The lattice constant monotonically decreases
from $a$ = 5.940 {\AA} for Li$_{1.1}$ZnAs to 5.929 {\AA} for $x$ =
0.10, as shown in Fig. 3(e), indicating the successful solid
solution of Cr for Zn up to the doping level of 10\%.

In Fig. 2, we show the electrical resistivity measured for
Li$_{1.1}$Zn$_{1-x}$Cr$_{x}$As with $x$ = 0, 0.03, 0.05, 0.10, 0.15.
For the parent semiconductor LiZnAs (note that Li concentration is
exactly 1), it displays a semiconducting behavior within the
temperature range of 2 K and 300 K \cite{Deng1}. 10$\%$ excess Li
readily changes the resistive behavior. As shown in Fig. 2, the
resistivity curve of Li$_{1.1}$ZnAs displays a metallic behavior
since excess Li provides additional carriers \cite{Deng1}.
Interestingly, once 3\% Cr is introduced, the resistivity turns to
monotonical increase toward lower temperature. This behavior is
conserved with the doping level up to 15 \%. The magnitudes of
resistivity at 4 K increase from 0.03 $\Omega$ mm of Li$_{1.1}$ZnAs,
to 0.3 $\Omega$ mm of the sample Li$_{1.1}$Zn$_{0.9}$Cr$_{0.1}$As,
and to 55 $\Omega$ mm of Li$_{1.1}$Zn$_{0.85}$Cr$_{0.15}$As. The
type of behavior in Cr doped specimens is likely arising from the
scattering of carriers by the magnetic fluctuations through exchange
interactions. We do not observe similar insulator to metal
transition that takes place at the doping level of $x$ = 0.03 for
(Ga$_{1-x}$Mn$_{x}$)As \cite{Jungwirth}.

We have also conducted the Hall effect measurements for the sample
of Li$_{1.1}$Zn$_{0.95}$Cr$_{0.05}$As at 200 K. Our results indicate
that the carriers are $p$-type, with a hole concentration of $p \sim
2 \times 10^{20}$ cm$^{-3}$. This carrier density is comparable to
that of Li$_{1.1}$Zn$_{1-x}$Mn$_{x}$As \cite{Deng1} but 3 orders
larger than that of Li$_{1.1}$Zn$_{1-x}$Mn$_{x}$P \cite{Deng2}. The
$p$-type carriers have also been observed for both
Li$_{1.1}$Zn$_{1-x}$Mn$_{x}$As and Li$_{1.04}$Zn$_{1-x}$Mn$_{x}$P,
which is in contrast with the intuitive expectation that excess Li
will render a $n$-type carriers. There are no convincing experiments
to clarify this issue so far. Based on first-principles
calculations, Deng et al \cite{Deng2} shows that excess Li$^{1+}$
ions are thermodynamically favored to occupy the Zn$^{2+}$ sites,
and each Li$^{1+}$ substitution for Zn$^{2+}$ will introduce a hole
carrier.

\begin{figure}
\includegraphics[width=8cm]{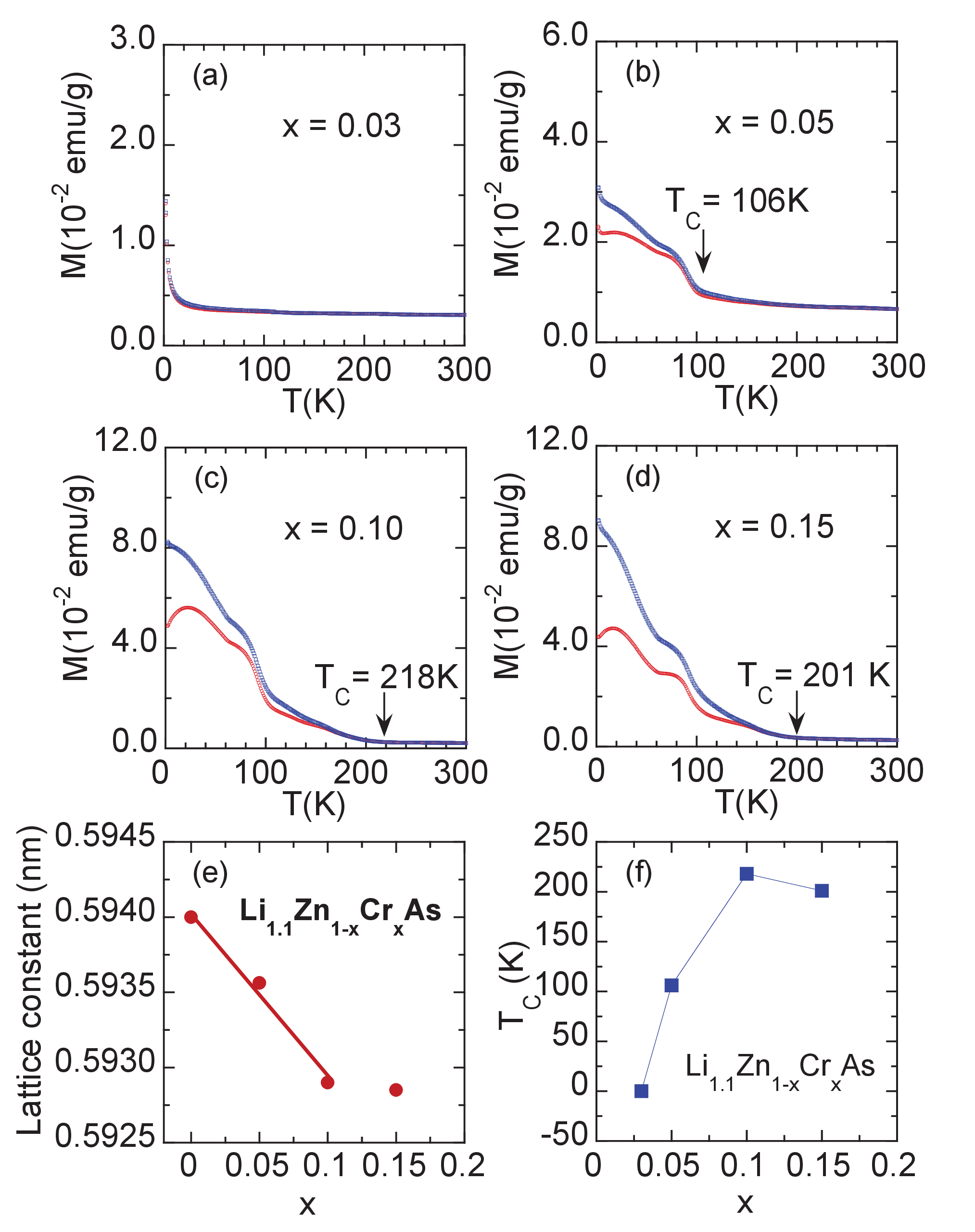}
\caption{\label{Fig3:epsart} (Color online)  (a)-(d) The
magnetization $M$ for Li$_{1.1}$Zn$_{1-x}$Cr$_{x}$As with $x$ =
0.03, 0.05, 0.10, 0.15 obtained in the zero field cooling (ZFC) and
field cooling (FC) in the external field of 1000 Oe; (e) The lattice
constant $a$ versus Cr concentration $x$, the solid line is guide
for eyes; (f) $T_C$ versus Cr concentration $x$.}
\end{figure}

In Fig. 3, we show the zero-field cooled (ZFC) and field cooled (FC)
measurements of the $dc$-magnetization $M$ of
Li$_{1.1}$Zn$_{1-x}$Cr$_{x}$As for $B_{ext}$ = 1000 Oe. For the
doping of $x$ = 0.03, we observe a strong increase of M at low
temperature, but no splitting is observed between ZFC and FC curves.
This indicates that although Cr substitution for Zn introduces local
moments, no magnetic ordering is formed. With the doping level
increasing to $x$ = 0.05, a significant increase in $M$ is observed
at the temperature of $\sim$ 106 K, and the ZFC and FC curves split,
indicating that the ferromagnetic ordering is taking place. $T_C$
increases to 218 K for the doping level of $x$ = 0.10, but decreases
to 201 K for $x$ = 0.15. We note that for $x$ = 0.10 and 0.15, two
successive steps at $\sim$ 100 K and $\sim$ 60 K have been observed
below $T_C$. This is possibly arising from either a spin state
change of Cr 3d electrons, or the inhomogeneous distribution of Cr
atoms in higher levels of Cr doped specimens. We are still working
to optimize the synthesis condition and to improve the sample
homogeneity. None the less, we believe that either reason does not
affect the fact that $T_C$ reaches $\sim$ 218 K at the average
doping level of 10\% Cr. We fit the temperature dependence of $M$
above $T_C$ to a Curie-Weiss law. The effective paramagnetic moment
is determined to be 2 $\sim$ 3$\mu _B$/Cr, which is smaller than the
case of 5.1 $\pm$ 0.4 $\mu _B$/ for (Ga,Cr)As \cite{Saito2},
indicating a likely different spin state in Li(Zn,Cr)As.

In Fig. 4, we show the isothermal magnetization of
Li$_{1.1}$Zn$_{1-x}$Cr$_{x}$As. For  $x$ = 0.03, no well formed
hysteresis loop is observed even at 2 K, which is consistent with
the absence of bifurcation of ZFC and FC curves. For $x$ = 0.05, a
parallelogram-shaped hysteresis loop with a coercive field of 497 Oe
is observed at 2 K. The coercive field decreases to 177 Oe at 100 K
and becomes zero at 300 K. For $x$ = 0.10 and 0.15, the coercive
fields at 2 K are 714 Oe and 764 Oe, respectively, which are larger
than $\sim$ 50 - 100 Oe of the same cubic structural
Li$_{1.1}$(Zn$_{0.97}$Mn$_{0.03}$)As \cite{Deng1},
Li$_{1.1}$(Zn$_{0.97}$Mn$_{0.03}$)P \cite{Deng2} and
(Ga$_{0.965}$Mn$_{0.035}$)As \cite{Ohno}.

\begin{figure}
\includegraphics[width=8cm]{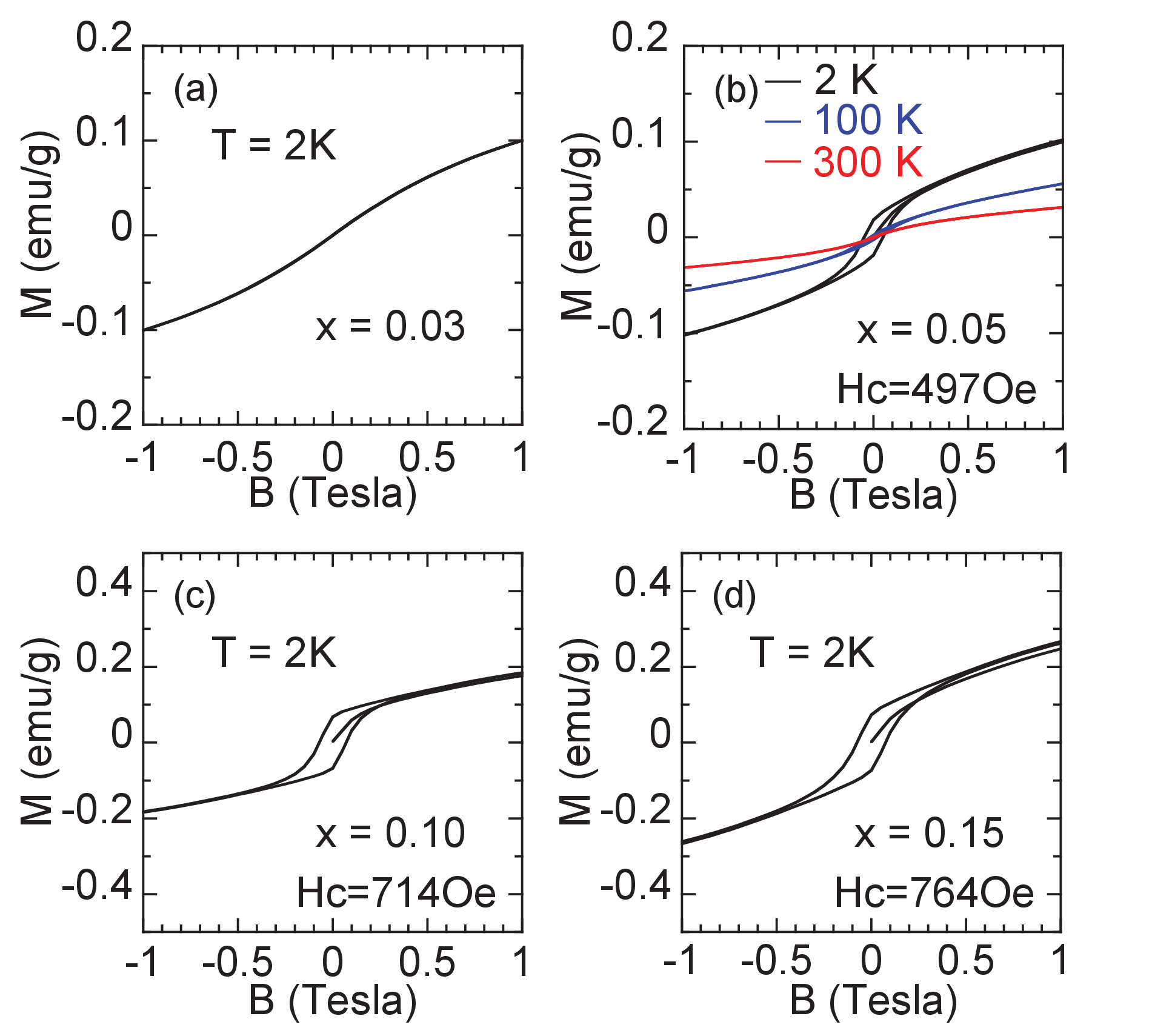}
\caption{\label{Fig4:epsart} (Color online) The isothermal
magnetization for Li$_{1.1}$Zn$_{1-x}$Cr$_{x}$As with $x$ = 0.03,
0.10, 0.15 measured at 2 K, and for $x$ = 0.05 measured at 2 K, 100
K and 300 K. $H_c$ values are the coercive filed at 2 K.}
\end{figure}

In diluted magnetic systems, the magnetic impurities can easily give
rise to spurious features of ``ferromagnetism", such as the
bifurcation of ZFC and FC curves and hysteresis loops \cite{Samarth,
Chambers}. For the $x$ = 0.05 specimen with $T_C$ = 106 K, the phase
is pure, as shown in Fig. 1(c). We detected tiny traces of CrAs
impurities for $x$ = 0.10 and 0.15 samples. However, CrAs is an
antiferromagnet with $T_N$ $\sim$ 260 K \cite{CrAs1, CrAs2}, which
does not likely contribute to the observed remanent magnetization in
these bulk form polycrystals. In the Li(Zn,Cr)As system, the
ordering temperature $T_f$ systematically increases with higher Cr
doping levels, along with the successful solid solution of Cr for Zn
as shown in Fig. 3(e), indicating that the magnetic ordering is
truly arising from the Cr atoms that substituted for the Zn atoms in
ionic sites. We would observe a single transition temperature for
all doping levels if the magnetic ordering arises from the same type
of magnetic impurity source or the uncompensated spins as observed
in thin films \cite{Dietl3}

In general, the bifurcation of ZFC and FC curves and the hysteresis
loops can be found not only in regular ferromagnets \cite{Ashcroft}
but also in spin glasses \cite{Fischer}. One decisive technique to
distinguish the two cases is neutron scattering, which can resolve
spatial spin correlations. We have conducted neutron diffraction
experiment on (Ba$_{0.7}$K$_{0.3}$)(Zn$_{0.9}$Mn$_{0.1}$)$_2$As$_2$
\cite{ZhaoK} and (La$_{0.9}$Sr$_{0.1}$)(Zn$_{0.9}$Mn$_{0.1}$)AsO
\cite{Lu}polycrystalline DMS specimens. Unfortunately, it is still
difficult to decouple the magnetic and structural Bragg peaks even
at 6 K due to the spatially dilute Mn moments. We are making efforts
to grow single crystals for high resolution neutron scattering
experiments at this stage.

In summary, we reported the synthesis and characterization of bulk
form diluted magnetic semiconductors Li(Zn,Cr)As with the ordering
temperature as high as $\sim$ 218 K. It is the first time that Cr
atoms are doped and a ferromagnetic ordering is observed in a I-II-V
semiconductor. The ferromagnetic ordering is arising from the Cr
atoms that substituted for Zn in ionic sites, and the Cr spins are
mediated by the carriers from excess 10 \% Li. Li(Zn,Cr)As therefore
represents a Cr induced DMS family that has the advantage of
decoupling of carrier and spins, whose spins are introduced by Cr
substitution for Zn and carriers are created by adding extra Li. Our
fabrication of new Li(Zn,Cr)As DMSs provides another DMS system for
the future research of precise controlling carriers/spins and their
individual influence to the ferromagnetic ordering.

The work at Zhejiang University was supported by National Basic
Research Program of China (No. 2011CBA00103, No. 2014CB921203), NSFC
(No. 11274268); F.L. Ning acknowledges partial support by the US NSF
PIRE (Partnership for International Research and Education:
OISE-0968226) and helpful discussion with J.H. Zhao, J.H. Dai, C.
Cao, C.Q. Jin and Y.J. Uemura.



\begin{thebibliography}{00}
\bibitem{Ohno}  H. Ohno, A. Shen, F. Matsukura, A. Oiwa, A. Endo, S. Katsumoto, and Y. Iye, Appl. Phys. Lett. \textbf{69}, 363 (1996).
\bibitem{Jungwirth}  T. Jungwirth, J. Sinova, J. Masek, J. Kucera, and A.H. MacDonald, Rev. Mod. Phys. \textbf{78}, 809 (2006).
\bibitem{Dietl1}  T. Dietl, Nature Materials  \textbf{9}, 965 (2010).
\bibitem{Zutic}  I. Zutic, J. Fabian, and S. Das Sarma, Rev. Mod. Phys.  \textbf{76}, 323 (2004).
\bibitem{ZhaoJH1}  L. Chen, S. Yan, P.F. Xu, W.Z. Wang, J.J. Deng, X. Qian, Y. Ji, J.H. Zhao, Appl. Phys. Lett. \textbf{95}, 182505 (2009).
\bibitem{ZhaoJH2}  L. Chen, X. Yang, F.H. Yang, J.H. Zhao, J. Misuraca, P. Xiong and S.V. Molnar, Nano Lett.  \textbf{11}, 2584 (2011).
\bibitem{Pajaczkowska}  A. Pajaczkowska, Prog. Cryst. Growth Charact. \textbf{1}, 289 (1978).
\bibitem{Furdyna}  J.K. Furdyna, J. Appl. Phys.  \textbf{64}, R29 (1988).
\bibitem{Wojtowicz}  T. Wojtowicz, T. Dietl, M. Sawicki, W. Plesiewicz, and J. Jaroszynski, Phys. Rev. Lett. \textbf{56}, 2419 (1986).
\bibitem{Morkoc} H. Morkoc, S. Strite, G. B. Gao, M. E. Lin, B. Sverdlov, and M. Burns, J. Appl. Phys.  \textbf{76}, 1363 (1994).
\bibitem{Shand} P.M. Shand, A.D. Christianson, T.M. Pekarek, L.S. Martinson, J.W. Schweitzer, I. Miotkowski, and B.C. Crooker, Phys. Rev. B  \textbf{58}, 12876 (1998).
\bibitem{Masek} J. Masek, J. Kudrnovsky, F. Maca, B.L. Gallagher, R.P. Campion, D.H. Gregory, and T. Jungwirth, Phys. Rev. Lett.  \textbf{98}, 067202 (2007).
\bibitem{Bacewicz} R.Bacewicz, and T.F. Ciszek,  Appl. Phys. Lett.  \textbf{52}, 1150 (1988).
\bibitem{Deng1}  Z. Deng, C.Q. Jin, Q.Q. Liu, X.C. Wang, J.L. Zhu, S.M. Feng, L.C. Chen, R.C. Yu, C. Arguello, T. Goko, F.L. Ning, J.S. Zhang, Y.Y. Wang, A.A. Aczel, T. Munsie, T.J. Williams, G.M. Luke, T. Kakeshita, S. Uchida, W. Higemoto, T.U. Ito, B. Gu, S. Maekawa, G.D. Morris and Y.J. Uemura, Nature Communications  \textbf{2}, 422 (2011).
\bibitem{Deng2} Z. Deng, K. Zhao, B. Gu, W. Han, J.L. Zhu, X.C. Wang, X. Li, Q.Q. Liu, R.C. Yu, T. Goko, B. Frandsen, L. Liu, J.S. Zhang, Y.Y. Wang, F.L. Ning, S. Maekawa, Y.J. Uemura and C.Q. Jin, Phys. Rev. B  \textbf{88}, 081203(R) (2013).
\bibitem{Ding2}  C. Ding, C. Qin,H.Y.  Man, T. Imai and F.L. Ning, Phys. Rev. B  \textbf{88}, 041108(R) (2013).
\bibitem{Ding1}  C. Ding, H.Y. Man, C.Qin, J.C. Lu, Y.L. Sun, Q. Wang, B.Q. Yu, C.M. Feng, T. Goko, C.J. Arguello, L. Liu, B.A. Frandsen, Y.J. Uemura, H.D. Wang, H. Luetkens, E. Morenzoni, W. Han, C.Q. Jin, T. Munsie, T.J. Williams, R.M. D¡¯Ortenzio, T. Medina, G.M. Luke, T. Imai, F.L. Ning, Phys. Rev. B  \textbf{88}, 041102(R) (2013).
\bibitem{Yangxj} X.J. Yang, Y.K. Li, C.Y. Shen, B.Q. Si, Y.L. Sun, Q. Tao, G.H. G.H. Cao, Z.A. Xu, and F.C. Zhang, Appl. Phys. Lett.  \textbf{103}, 022410 (2013).
\bibitem{ZhaoK} K. Zhao, Z. Deng, X.C. Wang, W. Han , J.L. Zhu, x. Li, Q.Q. Liu, R.C. Yu, T. Goko, B. Frandsen, L. Liu, F.L. Ning, Y.J. Uemura, H. Dabkowska, G.M. Luke, H. Luetkens, E. Morenzoni, S.R. Dunsiger, A. Senyshyn, P. B\"{o}ni and C.Q. Jin, Nature Communications  \textbf{4}, 1442  (2013).
\bibitem{Saito1} H. Saito, V. Zayets, S. Yamagata and K. Ando, Phys. Rev. Lett.  \textbf{90}, 207202  (2003).
\bibitem{ZhaoJH3} J. Lu, H.J. Meng, K. Zhu, L. Chen, P.F. Xu  Z. Xie and J.H. Zhao, Europhysics Letters  \textbf{89}, 67003 (2010).
\bibitem{ZhaoJH4} H. Wu, H.D. Gan, H.Z. Zheng, J. Lu, H. Zhu, Y. Ji  G.R. Li and J.H. Zhao, Solid State Communications \textbf{151}, 456 (2011).
\bibitem{Saito2} H. Saito, W. Zaets, R. Akimoto, K. Ando, Y. Mishima and M. Tanaka, J. Appl. Physics \textbf{89}, 7392 (2001).
\bibitem{Kuroda} S. Kuroda, N. Nishizawa, K. Takita, M. Mitome, Y. Bando, K. Osuch  and T. Dietla, Nature Materials \textbf{6}, 440 (2007).
\bibitem{CrAs1} K. Selte, A. Kjekshus,  W.E. Jamison A.F. Andresen and J. Engebretsen, Acta Chem. Scand. \textbf{25}, 1703 (1971).
\bibitem{CrAs2} A. Selte, H. Boller, and E.F. Bertaut, J. Phys. Chem. Solids \textbf{35}, 1139 (1974).
\bibitem{Samarth} N.Samarth, Nature Materials \textbf{9}, 955 (2010).
\bibitem{Chambers} S.Chambers, Nature Materials \textbf{9}, 956 (2010).
\bibitem{Dietl3} T. Dietl, T. Andrearczyk, A. Lipinsa, M. Kiecana, Maureen Tay, and Yihong Wu, Phys. Rev. B  \textbf{76}, 155312 (2007).
\bibitem{Ashcroft} N.W. Ashcroft and N.D. Mermin, Solid State Physics  \textbf {Holt, Rinehart and Winston, 1976}.
\bibitem{Fischer} K.H. Fischer and J.A. Hertz,   \textbf {Spin Glasses Cambridge University Press}, 1991.
\bibitem{Lu} J.C. Lu, H.Y. Man, C. Ding, Q. Wang, B.Q. Yu, S.L. Guo, Y.J. Uemura, W. Han, C.Q. Jin, H.D. Wang, B. Chen and F.L. Ning, Europhysics Letters  \textbf{103}, 67011 (2013).
\end{thebibliography}

\end{document}